\documentclass[aps,prb,twocolumn,amssymb,amsmath,showpacs,superscriptaddress]{revtex4-1}
\usepackage[T1]{fontenc}
\usepackage[latin9]{inputenc}
\usepackage{amsmath}
\usepackage{graphicx}
\usepackage{amssymb}
\usepackage{esint}
\usepackage{SIunits}
\begin{document}

\title{Low-Energy Effective Hamiltonian for Giant-Gap Quantum Spin Hall Insulators in Honeycomb X-Hydride/Halide (\textit{X} = N-Bi) Monolayers}

\author{Cheng-Cheng Liu}
%\email{ccliu@bit.edu.cn}
\affiliation {School of Physics, Beijing Institute of Technology, Beijing 100081, China}

\author{Shan Guan}
\affiliation {School of Physics, Beijing Institute of Technology, Beijing 100081, China}

\author{Zhigang Song}
\affiliation {State Key Laboratory for Mesoscopic Physics,  and School of Physics, Peking University, Beijing 100871, China}

\author{Shengyuan A. Yang}
\affiliation {Engineering Product Development, Singapore University of Technology and Design, Singapore 138682, Singapore}

\author{Jinbo Yang}
\affiliation {State Key Laboratory for Mesoscopic Physics,  and School of Physics, Peking University, Beijing 100871, China}
\affiliation {Collaborative Innovation Center of Quantum Matter, Beijing, China}

\author{Yugui Yao}
\email{ygyao@bit.edu.cn}
\affiliation {School of Physics, Beijing Institute of Technology, Beijing 100081, China}

\begin{abstract}
Using the tight-binding method in combination with first-principles calculations, we systematically derive a low-energy effective Hilbert subspace and Hamiltonian with spin-orbit coupling for two-dimensional hydrogenated and halogenated group-V monolayers. These materials are proposed to be giant-gap quantum spin Hall insulators with record huge bulk band gaps opened by the spin-orbit coupling at the Dirac points, e.g., from 0.74 to 1.08 eV in Bi\textit{X} (\textit{X} = H, F, Cl, and Br) monolayers. We find that the low-energy Hilbert subspace mainly consists of $p_{x}$ and $p_{y}$ orbitals from the group-V elements, and the giant first-order effective intrinsic spin-orbit coupling is from the on-site spin-orbit interaction. These features are quite distinct from those of group-IV monolayers such as graphene and silicene. There, the relevant orbital is $p_z$ and the effective intrinsic spin-orbit coupling is from the next-nearest-neighbor spin-orbit interaction processes. These systems represent the first real 2D honeycomb lattice materials in which the low-energy physics is associated with $p_{x}$ and $p_{y}$ orbitals. A spinful lattice Hamiltonian with an on-site spin-orbit coupling term is also derived, which could facilitate further investigations of these intriguing topological materials.
\end{abstract}

\pacs{73.43.-f, 73.22.-f, 71.70.Ej, 85.75.-d}

\maketitle

\section{INTRODUCTION}

\begin{figure}
\includegraphics[width=3.5in]{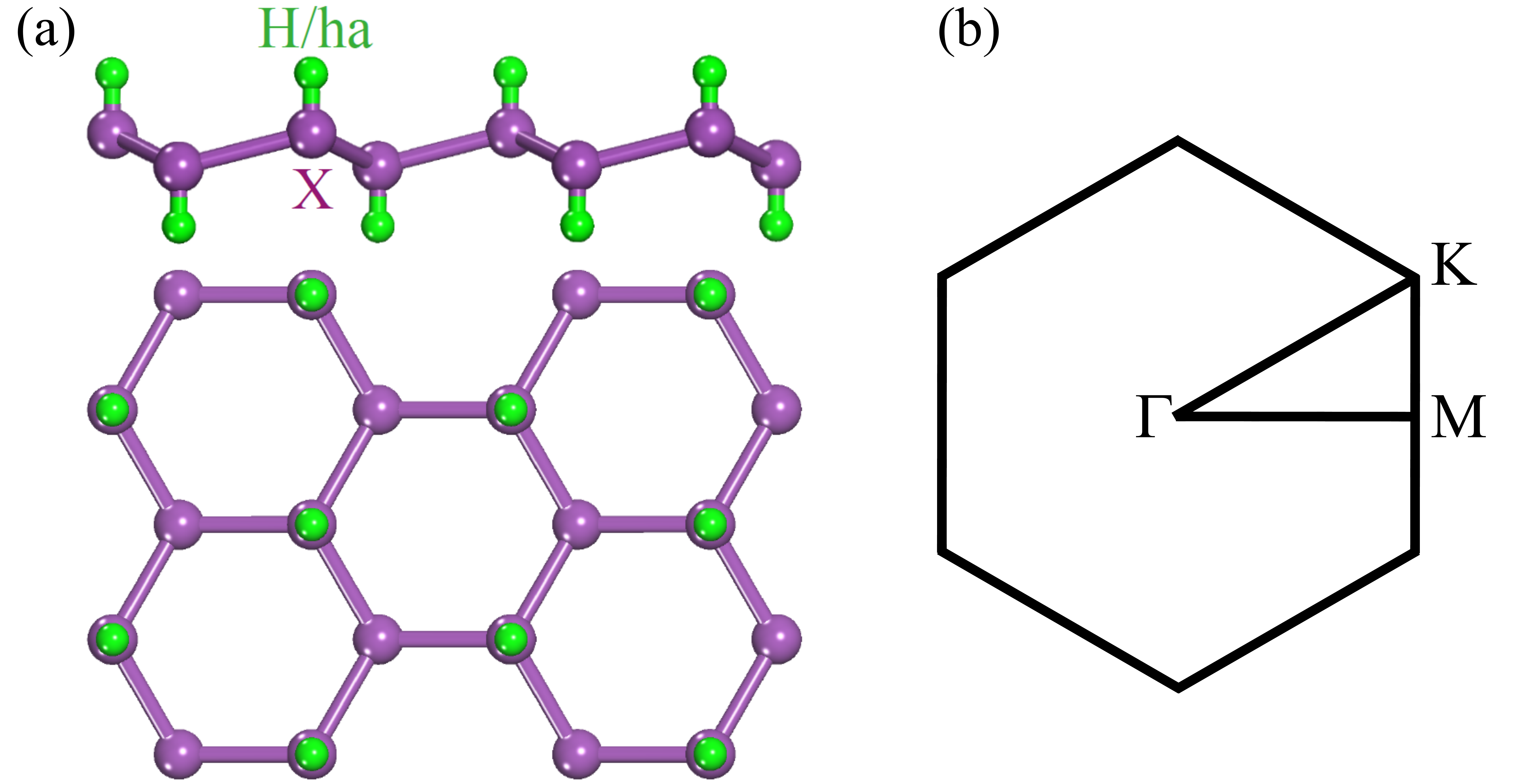}
\caption{(Color online). (a) The lattice geometry for 2D \textit{X}-hydride/halide (\textit{X} = N-Bi) monolayer from the side view (top) and top view (bottom). Note that two sets of sublattice in the honeycomb group V element \textit{X} are not coplanar (a buckled structure). The monolayer is alternatively hydrogenated or halogenated from both sides. (b) The first Brillouin zone of 2D \textit{X}-hydride/halide monolayer and the points of high symmetry.}\label{fig:geometry}
\end{figure}

Recent years have witnessed great interest in two-dimensional (2D) layered materials with honeycomb lattice structures. Especially, the 2D group-IV honeycomb lattice materials, such as successively fabricated graphene,~\cite{Gelm2007,RMP.Neto.2009} and silicene,~\cite{vogt_silicene:_2012,chen_evidence_2012} have attracted considerable attention both theoretically and experimentally due to their low-energy Dirac fermion behavior and promising applications in electronics. Recently, we have discovered stable 2D hydrogenated and halogenated group-V honeycomb lattices via first-principles (FP) calculations.~\cite{Song2014} Their structures are similar to that of a hydrogenated silicene (silicane), as shown in Fig.~\ref{fig:geometry}(a). In the absence of spin-orbit coupling (SOC), the band structures show linear energy crossing at the Fermi level around $K$ and $K'$ points of the hexagonal Brillouin zone. It is quite unusual that the low-energy bands of these materials are of $p_x$ and $p_y$ orbital character. Previous studies in the context of cold atoms systems have shown that $p_{x}$ and $p_{y}$ orbital character could lead to various charge and orbital ordered states as well as topological effects.~\cite{Wu2007,Wu2008} Our proposed materials, being the first real condensed matter systems in which the low-energy physics is associated with $p_{x}$ and $p_{y}$ orbitals, are therefore expected to exhibit rich and interesting physical phenomena.

The quantum spin Hall (QSH) insulator state has generated great interest in condensed matter physics and material science due to its scientific importance as a novel quantum state and its potential technological applications ranging from spintronics to topological quantum computation.~\cite{Hasan_2010,Qi_2010,Yan_2012} This novel electronic state is gaped in the bulk and conducts charge and spin in gapless edge states without dissipation protected by time-reversal symmetry. The concept of QSH effect was first proposed by Kane and Mele in graphene in which SOC opens a nontrivial band gap at the Dirac points.~\cite{kane2005a,kane2005b} Subsequent works, however, showed that the SOC for graphene is tiny, hence the effect is difficult to be detected experimentally.~\cite{yao2007,PhysRevB.Min.2006,Konschuh2010}
So far, QSH effect has only been demonstrated in HgTe/CdTe quantum wells,~\cite{Science.Bernevig.2006,Science.318.766} and experimental evidence for helical edge modes has been presented for inverted InAs-GaSb quantum wells.~\cite{PhysRevLett.Liu.2008,PRL.Knez.2011,PRL.Knez.2012} Nevertheless, these existing systems more or less have serious limitations like toxicity, difficulty in processing, and small bulk gap opened by SOC. Therefore, an easy and environmental friendly realization of a QSH insulator is much desired. Extensive effort has been devoted to the search for new QSH insulators with large SOC gap.~\cite{PhysRevLett.Murakami.2006,Liu2011,Hirahara2011,liu_quantum_2011,liu_low-energy_2011,PRX.Weeks.2011,Xu2013,Wang2013} For instance, new layered honeycomb lattice type materials such as silicene, germanene~\cite{liu_quantum_2011} or stanene~\cite{liu_low-energy_2011}, and chemically modified stanene~\cite{Xu2013} have been proposed.  Ultrathin Bi(111) films have drawn attention as a candidate QSH insulator, whose 2D topological properties have been reported.~\cite{Yang2012} An approach to design a large-gap QSH state on a semiconductor surface by a substrate orbital filtering process was also proposed.~\cite{Zhou2014}
However, desirable QSH insulators preferably with huge bulk gaps are still rare. A sizable bulk band gap in QSH insulators is essential for realizing many exotic phenomena and for fabricating new quantum devices that can operate at room temperature.

Using FP method, we have recently demonstrated that the QSH effect can be realized in the 2D hydrogenated and halogenated group-V honeycomb monolayers family, with a huge gap opened at the Dirac points due to SOC.~\cite{Song2014} Although the low-energy spectrum of these materials is similar to the 2D group-IV honeycomb monolayers such as graphene and silicene, the low-energy Hilbert space changes from the $p_z$ orbital to orbitals mainly consisting of $p_{x}$ and $p_{y}$ from the group-V atoms (N-Bi). Moreover, the nature of the effective SOC differs between the two systems. Motivated by the fundamental interest associated with the QSH effect and huge SOC gaps in these novel 2D materials, we develop a low-energy effective model Hamiltonian that captures their essential physics.
In addition, we propose a minimal four-band lattice Hamiltonian with the on-site SOC term using only the $p_x$ and $p_y$ orbitals.

\begin{figure*}
\includegraphics[width=7in]{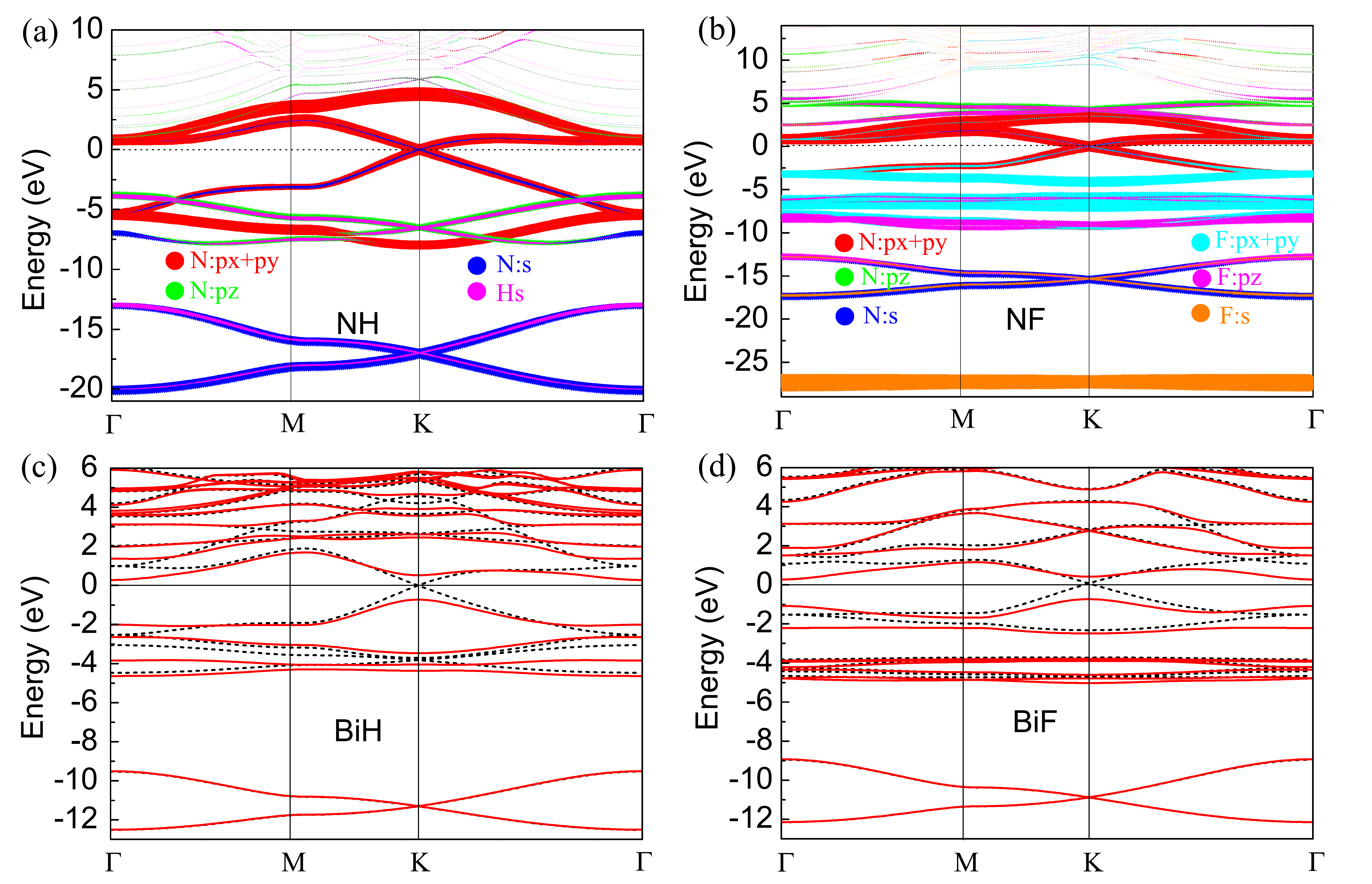}
\caption{(Color online). (a)(b) The partial band structure projection for NH and NF without SOC, respectively. Symbol size is proportional to the population in the corresponding states. The Fermi level is indicated by the dotted line. (c)(d) Band structures for BiH and BiF without (black dash lines) and with (red solid lines) SOC. The four band structures are obtained from the first-principles methods implemented in the VASP package~\cite{Kresse1996} using projector augmented wave pseudo-potential, and the exchange-correlation is treated by PAW-GGA. The Fermi level is indicated by the solid line.}\label{fig:Bands}
\end{figure*}

From the symmetry analysis, the next-nearest-neighbor (NNN) intrinsic Rashba SOC should exist in these systems due to the low-buckled structure, similar to the case of silicene.~\cite{liu_low-energy_2011} However, as we shall see, the dominant effect is from the much larger first-order SOC of on-site origin. Therefore, in the following discussion, we shall focus on the first-order on-site SOC and neglect the higher-order effects. This point will be further discussed later in this paper.

The paper is organized as follows. In Sec. II, we derive step by step the low-energy effective Hilbert subspace and Hamiltonian for honeycomb \textit{X}-hydride (\textit{X} = N-Bi) monolayers, and also investigate in detail the effective SOC. Section III presents the derivation of the low-energy effective model for \textit{X}-halide (\textit{X} = N-Bi, halide=F-I) honeycomb monolayers.
In Sec. IV, a simple spinful lattice Hamiltonian for the honeycomb \textit{X}-hydride/halide monolayers family is constructed. We conclude in Sec. V with a brief discussion of the effective SOC and present a summary of our results.

\section{Low-energy effective Hamiltonian for honeycomb \textit{X}H(\textit{X} = N-Bi) monolayers}

\subsection{Low-energy Hilbert subspace and effective Hamiltonian without SOC}

As is shown in Fig.~\ref{fig:geometry}(a), there are two distinct sites A and B in the unit cell of \textit{X}-hydride (\textit{X} = N-Bi) honeycomb lattice with full hydrogenation from both sides of the 2D \textit{X} honeycomb sheet. The primitive lattice vectors are chosen as
$\vec{a}_{1}=a(1/2,\sqrt{3}/2)$ and $\vec{a}_{2}=a(-1/2,\sqrt{3}/2)$, where $a$ is the lattice constant. We consider the outer shell orbitals of textit{X} (\textit{X} = N-Bi), namely $s$, $p_{x}$, $p_{y}$, $p_{z}$, and also the $s$ orbital of H in the modeling. Therefore, in the representation $\{|p_{y}^{A}\rangle,|p_{x}^{A}\rangle,|p_{z}^{A}\rangle,|s_{H}^{A}\rangle,|s^{A}\rangle,|p_{y}^{B}\rangle,|p_{x}^{B}\rangle,|p_{z}^{B}\rangle,|s_{H}^{B}\rangle,|s^{B}\rangle\}$  (for simplicity, the Dirac ket symbol is omitted in the following), the Hamiltonian (without SOC) at $K$ point with the nearest-neighbor hopping considered in the Slater-Koster formalism~\cite{Slater1954} reads

\begin{eqnarray}\label{H0}
H_{0}=\left(\begin{array}{cc}
H_{0}^{AA} & H_{0}^{AB}\\
H_{0}^{AB\dagger} & H_{0}^{BB}
\end{array}\right),
\end{eqnarray}
with
\begin{eqnarray}
H_{0}^{AA}=\left[\begin{array}{ccccc}
0 & 0 & 0 & 0 & 0\\
0 & 0 & 0 & 0 & 0\\
0 & 0 & 0 & -V_{sp\sigma}^{H} & 0\\
0 & 0 & -V_{sp\sigma}^{H} & \Delta_{H} & V_{ss\sigma}^{H}\\
0 & 0 & 0 & V_{ss\sigma}^{H} & \Delta
\end{array}\right],
\end{eqnarray}
\begin{eqnarray}
H_{0}^{AB}=\left[\begin{array}{ccccc}
-V_{1}^{'} & -iV_{1}^{'} & 0 & 0 & V_{2}^{'}\\
-iV_{1}^{'} & V_{1}^{'} & 0 & 0 & -iV_{2}^{'}\\
0 & 0 & 0 & 0 & 0\\
0 & 0 & 0 & 0 & 0\\
-V_{2}^{'} & iV_{2}^{'} & 0 & 0 & 0
\end{array}\right],
\end{eqnarray}
\begin{eqnarray}
H_{0}^{BB}=\left[\begin{array}{ccccc}
0 & 0 & 0 & 0 & 0\\
0 & 0 & 0 & 0 & 0\\
0 & 0 & 0 & V_{sp\sigma}^{H} & 0\\
0 & 0 & V_{sp\sigma}^{H} & \Delta_{H} & V_{ss\sigma}^{H}\\
0 & 0 & 0 & V_{ss\sigma}^{H} & \Delta
\end{array}\right],
\end{eqnarray}
where $V_{sp\sigma}^{H}$ ($V_{ss\sigma}^{H}$) is the hopping between the $p_z$ ($s$) orbital from \textit{X} atom and the $s$ orbital from H, and $V_{1}^{'}\equiv(3/4)\left(V_{pp\pi}-V_{pp\sigma}\right)$ and $V_{2}^{'}\equiv(3/2)V_{sp\sigma}$ with $V_{pp\pi}$, $V_{sp\sigma}$, and $V_{pp\sigma}$ being the standard Slater-Koster hopping parameters. $\Delta$ and $\Delta_{H}$ are on-site energies for $s$ orbitals of atom \textit{X} and of atom H, respectively. The on-site energies for $p$ orbitals are taken to be zero.

To diagonalize the Hamiltonian, we first perform the following unitary transformation:
\begin{equation}\label{u4x4}
\begin{split} & \varphi_{1}^{A}=-\frac{1}{\sqrt{2}}\left(p_{x}^{A}+ip_{y}^{A}\right)=|p_{+}^{A}\rangle,\\
 & \varphi_{2}^{B}=\frac{1}{\sqrt{2}}\left(p_{x}^{B}-ip_{y}^{B}\right)=|p_{-}^{B}\rangle,\\
 & \varphi_{3}=\frac{1}{\sqrt{2}}\left[-\frac{1}{\sqrt{2}}\left(p_{x}^{A}-ip_{y}^{A}\right)-\frac{1}{\sqrt{2}}\left(p_{x}^{B}+ip_{y}^{B}\right)\right],\\
 & \varphi_{4}=\frac{1}{\sqrt{2}}\left[\frac{1}{\sqrt{2}}\left(p_{x}^{A}-ip_{y}^{A}\right)-\frac{1}{\sqrt{2}}\left(p_{x}^{B}+ip_{y}^{B}\right)\right].
\end{split}
\end{equation}

In the basis
$\{\varphi_{1}^{A},s^{B},s_{H}^{B},p_{z}^{B},\varphi_{2}^{B},s^{A},s_{H}^{A},p_{z}^{A},\varphi_{3},\varphi_{4}\}$, the Hamiltonian can be written as a block-diagonal form with three decoupled blocks $H_{\alpha}$, $H_{\beta}$, and $H_{\gamma}$:
\begin{equation}\label{H2}
H_{0}\longrightarrow H_{1}=U_{1}^{\dagger}H_{0}U_{1},
\end{equation}
\begin{eqnarray}\label{U1}
U_{1}=\left[\begin{array}{cccccccccc}
\frac{-i}{\sqrt{2}} & 0 & \frac{i}{2} & \frac{-i}{2} & 0 & 0 & 0 & 0 & 0 & 0\\
\frac{-1}{\sqrt{2}} & 0 & \frac{-1}{2} & \frac{1}{2} & 0 & 0 & 0 & 0 & 0 & 0\\
0 & 0 & 0 & 0 & 1 & 0 & 0 & 0 & 0 & 0\\
0 & 0 & 0 & 0 & 0 & 1 & 0 & 0 & 0 & 0\\
0 & 0 & 0 & 0 & 0 & 0 & 1 & 0 & 0 & 0\\
0 & \frac{-i}{\sqrt{2}} & \frac{-i}{2} & \frac{-i}{2} & 0 & 0 & 0 & 0 & 0 & 0\\
0 & \frac{1}{\sqrt{2}} & \frac{-1}{2} & \frac{-1}{2} & 0 & 0 & 0 & 0 & 0 & 0\\
0 & 0 & 0 & 0 & 0 & 0 & 0 & 1 & 0 & 0\\
0 & 0 & 0 & 0 & 0 & 0 & 0 & 0 & 1 & 0\\
0 & 0 & 0 & 0 & 0 & 0 & 0 & 0 & 0 & 1
\end{array}\right],
\end{eqnarray}
\begin{eqnarray}\label{H1}
H_{1}=H_{\alpha}\oplus H_{\beta}\oplus H_{\gamma},
\end{eqnarray}
with
\begin{eqnarray}\label{Ha}
H_{\alpha}=\left[\begin{array}{cccc}
0 & iV_{2} & 0 & 0\\
-iV_{2} & \Delta & V_{ss\sigma}^{H} & 0\\
0 & V_{ss\sigma}^{H} & \Delta_{H} & V_{sp\sigma}^{H}\\
0 & 0 & V_{sp\sigma}^{H} & 0
\end{array}\right],
\end{eqnarray}
\begin{eqnarray}\label{Hb}
H_{\beta}=\left[\begin{array}{cccc}
0 & -iV_{2} & 0 & 0\\
iV_{2} & \Delta & V_{ss\sigma}^{H} & 0\\
0 & V_{ss\sigma}^{H} & \Delta_{H} & -V_{sp\sigma}^{H}\\
0 & 0 & -V_{sp\sigma}^{H} & 0
\end{array}\right],
\end{eqnarray}
\begin{eqnarray}
H_{\gamma}=\text{diag}\left\{ V_{1},-V_{1}\right\},
\end{eqnarray}
where $V_{1}=2V_{1}^{'}$ and $V_{2}=\sqrt{2}V_{2}^{'}$.

The eigenvectors for the first diagonal block $H_{\alpha}$ can be easily obtained as
\begin{eqnarray}\label{Ha-vec}
|\varepsilon_{i}\rangle=\frac{1}{N_{i}}\left[\begin{array}{c}
1\\
-i\frac{\varepsilon_{i}}{V_{2}}\\
-i\frac{\varepsilon_{i}^{2}-\Delta\varepsilon_{i}-V_{2}^{2}}{V_{2}V_{ss\sigma}^{H}}\\
-i\frac{V_{sp\sigma}^{H}}{\varepsilon_{i}}\frac{\varepsilon_{i}^{2}-\Delta\varepsilon_{i}-V_{2}^{2}}{V_{2}V_{ss\sigma}^{H}}
\end{array}\right],
\end{eqnarray}
where $\varepsilon_{i}$  and $N_{i}$ $(i=1,2,3,4)$  are the corresponding eigenvalues and normalization factors, respectively. Therefore, upon performing the unitary transformation $\{ \phi_{1},\phi_{2},\phi_{3},\phi_{4}\} =\{ \varphi_{1}^{A},s^{B},s_{H}^{B},p_{z}^{B}\} U_{\alpha}$  with $U_{\alpha}=\left\{ |\varepsilon_{i}\rangle\right\} _{i=1,2,3,4}\equiv\{ u_{ji}^{\alpha}\}$, the above upper-left $4\times4$ block $H_\alpha$ is diagonalized.

For the second diagonal block $H_{\beta}$, its eigenvalues are denoted as $\varepsilon_{4+i}$ $(i=1,2,3,4)$, and it can be easily shown that $\varepsilon_{4+i}=\varepsilon_{i}$, where $\varepsilon_{i}$  are eigenvalues of $H_{\alpha}$. This
is consistent with FP results, i.e., there are four two-fold degeneracy points at $K$ point as shown in Fig.~\ref{fig:Bands}(a).
The eigenvectors of $H_{\beta}$ are given by
\begin{eqnarray}\label{Hb-vec}
|\varepsilon_{i}\rangle=\frac{1}{N_{i}}\left[\begin{array}{c}
1\\
i\frac{\varepsilon_{i}}{V_{2}}\\
i\frac{\varepsilon_{i}^{2}-\Delta\varepsilon_{i}-V_{2}^{2}}{V_{2}V_{ss\sigma}^{H}}\\
-i\frac{V_{sp\sigma}^{H}}{\varepsilon_{i}}\frac{\varepsilon_{i}^{2}-\Delta\varepsilon_{i}-V_{2}^{2}}{V_{2}V_{ss\sigma}^{H}}
\end{array}\right],
\end{eqnarray}
where $\varepsilon_{i}$  and $N_{i}$ $(i=5,6,7,8)$ are the corresponding eigenvalues and normalization factors. Similar to the case of $H_{\alpha}$, upon performing the unitary transformation $\{ \phi_{5},\phi_{6},\phi_{7},\phi_{8}\} =\{ \varphi_{2}^{B},s^{A},s_{H}^{A},p_{z}^{A}\} U_{\beta}$  with $U_{\beta}=\left\{ |\varepsilon_{i+4}\rangle\right\} _{i=1,2,3,4}\equiv\{ u_{ji}^{\beta}\}$, the block $H_{\beta}$ is diagonalized.

The third block $H_{\gamma}$ is already diagonal with eigenvalues $\{ V_{1},-V_{1}\}$ and eigenvectors $\{ \varphi_{3},\varphi_{4}\} \equiv\{ \phi_{9},\phi_{10}\}$. Therefore, in the new basis $\left\{ \phi_{1},\phi_{2},\phi_{3},\phi_{4},\phi_{5},\phi_{6},\phi_{7},\phi_{8},\phi_{9},\phi_{10}\right\} \equiv\left\{ \varphi_{1}^{A},s^{B},s_{H}^{B},p_{z}^{B},\varphi_{2}^{B},s^{A},s_{H}^{A},p_{z}^{A},\varphi_{3},\varphi_{4}\right\} U_{2}$, where $U_{2}\equiv u^{\alpha}\oplus u^{\beta}\oplus I_{2\times2}$, the total Hamiltonian ~(\ref{H0}) takes a fully diagonlized form. The whole diagonalization process can be summarized as follows:
\begin{equation}\label{Utrans}
\begin{split}
& \left\{ \phi_{1},\phi_{2},\phi_{3},\phi_{4},\phi_{5},\phi_{6},\phi_{7},\phi_{8},\phi_{9},\phi_{10}\right\} \\
& =\left\{p_{y}^{A},p_{x}^{A}\text{,}p_{z}^{A},s_{H}^{A},s^{A},p_{y}^{B},p_{x}^{B},p_{z}^{B},s_{H}^{B},s^{B}\right\}U, \\
\end{split}
\end{equation}
where
\begin{equation}\label{U}
U=U_{1}U_{2},
\end{equation}
\begin{equation}\label{U}
H_{0}\longrightarrow H_{0}^{'}=U^{\dagger}H_{0}U,
\end{equation}
\begin{equation}\label{H0'}
H_{0}^{'}=\text{diag}\left\{\varepsilon_{1},\varepsilon_{2},\varepsilon_{3},\varepsilon_{4},\varepsilon_{5},\varepsilon_{6},\varepsilon_{7},\varepsilon_{8},V_{1},-V_{1}\right\}.
\end{equation}

From the band components projection as shown in Fig.~\ref{fig:Bands}(a), in the vicinity of the Dirac points (around Fermi level), the main components of the band come from the  $p_{x}$  and $p_{y}$ orbitals of group-V element textit{X} mixed with a small amount of $s$ orbital of textit{X}. Compared with the expressions of the eigenstates obtained above, we find that the orbital features agree with that of $|\varepsilon_{1}\rangle$ and $|\varepsilon_{5}\rangle$ if we take their eigenenergies as the Fermi energy. Therefore the corresponding states $\phi_{1}$ and $\phi_{5}$ constitute the low-energy Hilbert subspace. In the following, we will give the explicit forms of the low-energy states $\phi_{1}$  and $\phi_{5}$ as well as their eigenvalues.

Note that, in the above $4\times4$  $H_{\alpha}$, the scale of the $2\times2$  non-diagonal block $H_{\alpha12}$ is smaller than the difference of the typical eigenvalues between the upper $2\times2$ diagonal block $H_{\alpha11}$ and the lower $2\times2$  diagonal block $H_{\alpha22}$.
Hence, through the downfolding procedure~\cite{winkler_spin-orbit_2003}, we could obtain the low-energy effective Hamiltonian as
\begin{equation}\label{2ndperturbation}
H_{\alpha11}^\text{eff}=H_{\alpha11}+H_{\alpha12}\left(\varepsilon-H_{\alpha22}\right)^{-1}H_{\alpha21}.
\end{equation}
Up to the second order, one obtains
\begin{equation}\label{E1}
\varepsilon_{1}=\frac{1}{2}\left(\Delta^{'}+\sqrt{\Delta^{'2}+4V_{2}^{2}}\right),
\end{equation}
with
\begin{equation}\label{E1-s}\begin{split}
\Delta^{'}=\Delta+\frac{\varepsilon V_{ss\sigma}^{H}{}^{2}}{\varepsilon^{2}-\Delta_{H}\varepsilon-V_{sp\sigma}^{H}{}^{2}},\\
\varepsilon=\frac{1}{2}\left(\Delta+\sqrt{\Delta^{2}+4V_{2}^{2}}\right).
\end{split}
\end{equation}
Consequently, we can obtain the explicit expressions of $|\varepsilon_{1}\rangle\equiv\left\{ u_{j1}^{\alpha}\right\} _{j=1,4}$  and $\phi_{1}$. In a similar way, the explicit expressions of $|\varepsilon_{5}\rangle\equiv\{ u_{j1}^{\beta}\} _{j=1,4}$  and $\phi_{5}$  can also be obtained. So far, we have obtained the eigenvalues $\varepsilon_{1}=\varepsilon_{5}$ [Eqs.~\eqref{E1} and ~\eqref{E1-s}] and the corresponding low-energy Hilbert subspace consisting of $\phi_{1}$ and $\phi_{5}$,
\begin{equation}\label{full-Hilbert}
\begin{split}
&\phi_{1}=u_{11}^{\alpha}\varphi_{1}^{A}+u_{21}^{\alpha}s^{B}+u_{31}^{\alpha}s_{H}^{B}+u_{41}^{\alpha}p_{z}^{B}, \\
&\phi_{5}=u_{11}^{\alpha}\varphi_{2}^{B}-u_{21}^{\alpha}s^{A}-u_{31}^{\alpha}s_{H}^{A}+u_{41}^{\alpha}p_{z}^{A}. \\
\end{split}
\end{equation}
The above coefficients $\{ u_{j1}^{\alpha}\} _{j=1,4}$ are given in Eq.~\eqref{Ha-vec}.

Further simplification could be made in order to capture the main physics. We can omit the second-order correction for the eigenvalues and the first-order correction for the eigenvectors, i.e., the terms $(u_{31}^{\alpha}s_{H}^{B}+u_{41}^{\alpha}p_{z}^{B})$  for $\phi_{1}$  and $(-u_{31}^{\alpha}s_{H}^{A}+u_{41}^{\alpha}p_{z}^{A})$  for $\phi_{5}$, and only keep the zeroth-order eigenvectors and eigenvalues,
\begin{equation}\label{full-Hilbert-simple-E}
\begin{split}
 & \phi_{1}=u_{11}^{\alpha}\varphi_{1}^{A}+u_{21}^{\alpha}s^{B}, \\
 & \phi_{5}=u_{11}^{\alpha}\varphi_{2}^{B}-u_{21}^{\alpha}s^{A}, \\
 & \varepsilon_{1}=\varepsilon=\frac{1}{2}\left(\Delta+\sqrt{\Delta^{2}+4V_{2}^{2}}\right).
\end{split}
\end{equation}
This approximation is justified by our FP calculations, namely in the vicinity of the Fermi level, $p_{x}$, $p_{y}$,  and $s$ orbitals overwhelmingly dominate over the $s_{H}$ and $p_{z}$ orbitals in the band components.

In the Hamiltonian ~\eqref{H0'}, one can take the Fermi energy $E_{F}=\varepsilon_{1}=\varepsilon_{5}$  as energy zero point. Hence, states $\phi_{1}$ and $\phi_{5}$, which constitute the low-energy Hilbert subspace, take the following explicit forms:
\begin{equation}\label{Ksimple-Hilbert}
\begin{split}
 & \phi_{1}=u_{11}^{\alpha}\left[-\frac{1}{\sqrt{2}}\left(p_{x}^{A}+ip_{y}^{A}\right)\right]+u_{21}^{\alpha}s^{B},  \\
 & \phi_{5}=u_{11}^{\alpha}\left[\frac{1}{\sqrt{2}}\left(p_{x}^{B}-ip_{y}^{B}\right)\right]-u_{21}^{\alpha}s^{A}, \\
\end{split}
\end{equation}
with
\begin{equation*}\label{}
\begin{split}
 & u_{11}^{\alpha}=\frac{\left(-\Delta+\sqrt{\Delta^{2}+18V_{sp\sigma}^{2}}\right)}{\sqrt{2\Delta^{2}+36V_{sp\sigma}^{2}-2\Delta\sqrt{\Delta^{2}+18V_{sp\sigma}^{2}}}},   \\
 & u_{21}^{\alpha}=\frac{-3\sqrt{2}iV_{sp\sigma}}{\sqrt{2\Delta^{2}+36V_{sp\sigma}^{2}-2\Delta\sqrt{\Delta^{2}+18V_{sp\sigma}^{2}}}}. \\
\end{split}
\end{equation*}

Since we are interested in the low-energy physics near the Dirac point, we perform the small $\vec{k}$ expansion around $K$  by $\vec{k}\rightarrow\vec{k}+K$ and keep the terms that are first order in $\vec{k}$. We find that
\begin{eqnarray}\label{HK}
H_{K}=\left(\begin{array}{cc}
0 & v_{F}k_{-}\\
v_{F}k_{+} & 0
\end{array}\right),
\end{eqnarray}
with $v_{F}$ being the Fermi velocity
\begin{equation}\label{vf}
v_{F}=\frac{\sqrt{3}a}{2}\left[\frac{1}{2}\mid u_{11}^{\alpha}\mid^{2}\left(V_{pp\sigma}-V_{pp\pi}\right)+\mid u_{21}^{\alpha}\mid^{2}V_{ss\sigma}\right],
\end{equation}
and
\begin{equation*}\label{}
k_\pm=k_{x}\pm ik_{y}.
\end{equation*}

Either following similar procedures, or using the inversion symmetry (or time-reversal symmetry ) of the system, we can easily obtain the low-energy Hilbert subspace and the low-energy effective Hamiltonian around the $K'$ point. Finally, we can summarize the basis for the low-energy Hilbert subspace as
\begin{equation}\label{simple-Hilbert}
\begin{split}
 & \phi_{1}=u_{11}^{\alpha}\left[-\frac{1}{\sqrt{2}}\left(p_{x}^{A}+i\tau_{z}p_{y}^{A}\right)\right]+u_{21}^{\alpha}\tau_{z}s^{B},  \\
 & \phi_{5}=u_{11}^{\alpha}\left[\frac{1}{\sqrt{2}}\left(p_{x}^{B}-i\tau_{z}p_{y}^{B}\right)\right]-u_{21}^{\alpha}\tau_{z}s^{A}, \\
\end{split}
\end{equation}
and the low-energy effective Hamiltonian without SOC reads
\begin{equation}\label{Heff}
H_{\tau}=v_{F}\left(k_{x}\sigma_{x}+\tau_{z}k_{y}\sigma_{y}\right),
\end{equation}
where Pauli matrices $\sigma$ denote the orbital basis degree of freedom, and $\tau_{z}=\pm 1$ labels the two valleys $K$ and $K'$. Note that under the space inversion operation $P=\sigma_{x}\tau_{x}$ and the time-reversal operation $T=\tau_{x}\hat{K}$  ($\hat{K}$ is the complex conjugation operator), the above low-energy effective Hamiltonian [Eq.~\eqref{Heff}] is invariant.

\subsection{Low-energy effective Hamiltonian involving SOC}

The SOC can be written as
\begin{equation}\label{SOC}
H_{so}=\xi_{0}\hat{L}\cdot\hat{s}=\frac{\xi_{0}}{2}\left(\frac{L_{+}s_{-}+L_{-}s_{+}}{2}+L_{z}s_{z}\right),
\end{equation}
where $s_{\pm}=s_{x}\pm is_{y}$ and $L_{\pm}=L_{x}\pm iL_{y}$  denote the ladder operators for the spin and orbital angular momenta, respectively. Here $\hat{s}=(\hbar/2)\vec{s}$, and in the following we shall take $\hbar=1$. $\xi_{0}$  is the magnitude of atomic SOC. Because of the presence of $p_x$ and $p_y$ orbital component in the low-energy Hilbert subspace [Eq.~\eqref{simple-Hilbert}] $\left\{ \phi_{1},\phi_{5}\right\} \otimes\{ \uparrow,\downarrow\}$, an on-site effective SOC is generated with
\begin{equation}\label{Hsoc}
H_{so}=\lambda_{so}\tau_{z}\sigma_{z}s_{z},
\end{equation}
where
\begin{equation}\label{lamdaso}
\begin{split}
\lambda_{so}
& =\frac{1}{2}\mid u_{11}^{\alpha}\mid^{2}\xi_{0} \\
&=\frac{1}{2}\left[1-\frac{9V_{sp\sigma}^{2}}{\Delta^{2}-\Delta\sqrt{\Delta^{2}+18V_{sp\sigma}^{2}}+18V_{sp\sigma}^{2}}\right]\xi_{0}. \end{split}
\end{equation}
Again we stress that in the honeycomb textit{X}-hydride monolayers the dominant intrinsic effective SOC is on-site rather than from the NNN hopping processes as in the original Kane-Mele model.

Consequently, from the above Hamiltonian ~(\ref{Heff}) and (\ref{Hsoc}), we obtain the generic low-energy effective Hamiltonian around the Dirac points acting on the low-energy Hilbert subspace:
\begin{equation}\label{Heffsoc}
H_\text{eff}=H_{\tau}+H_{so}=v_{F}\left(k_{x}\sigma_{x}+\tau_{z}k_{y}\sigma_{y}\right)+\lambda_{so}\tau_{z}\sigma_{z}s_{z},
\end{equation}
where the analytical expressions for Fermi velocity $v_{F}$ and magnitude of intrinsic effective SOC $\lambda_{so}$  are given in Eqs.~\eqref{vf} and~\eqref{lamdaso}, whose explicit values are presented in Table~\ref{tab:XH-parameter} via FP calculations. Again we note that the above spinful low-energy effective Hamiltonian is invariant under both the space-inversion symmetry operation and time-reversal symmetry operation with $T=is_y\tau_{x}\hat{K}$.

The two model parameters $v_F$ and $\lambda_{so}$ can be obtained by fitting the band dispersions of the FP results. Their values are listed in Table \ref{tab:XH-parameter}.

\begin{table}
\caption{Values of Fermi velocity $v_{F}$ and magnitude of intrinsic SOC $\lambda_{so}$ for textit{X}-hydride honeycomb monolayers obtained from FP calculations. Note that $\lambda_{so}=E_{g}/2$, with $E_{g}$ the gap opened by SOC at the Dirac point.}\label{tab:XH-parameter}
\begin{ruledtabular}
\begin{tabular}{ccccccc}
& system & $v_{F}\left(10^{5}m/s\right)$ & $\lambda_{so}\left(eV\right)$ \\
\hline
& NH & 6.8 & $6.7\times10^{-3}$ \\
& PH & 8.3 & $18\times10^{-3}$  \\
& AsH & 8.7 & $97\times10^{-3}$ \\
& SbH & 8.6 & 0.21 \\
& BiH & 8.9 & 0.62 \\
\end{tabular}
\end{ruledtabular}
\end{table}

\section{Low-energy effective Hamiltonian for honeycomb textit{X}-halide (\textit{X} = N-Bi) monolayers}

\subsection{Low-energy Hilbert subspace and effective Hamiltonian without SOC}

For the textit{X}-halide (\textit{X} = N-Bi) systems, the outer shell orbitals of X labeled as $Xs$, $Xp_{x}$, $Xp_{y}$, $Xp_{z}$, and the outer shell orbitals of halogen labeled as $Hs$, $Hp_{x}$, $Hp_{y}$, $Hp_{z}$  with (H=F-I) are taken into account in the following derivation. As is shown in Fig.~\ref{fig:geometry}(a), there are also two distinct sites A and B in the honeycomb lattice unit cell of textit{X}-halide with full halogenation from both sides of the 2D textit{X} honeycomb sheet. In the representation $\{ Xp_{y}^{A}$, $Xp_{x}^{A}$, $Xp_{z}^{A}$, $Hp_{z}^{A}$, $Hp_{y}^{A}$, $Hp_{x}^{A}$, $Hs^{A}$, $Xs^{A}$, $Xp_{y}^{B}$, $Xp_{x}^{B}$, $Xp_{z}^{B}$, $Hp_{z}^{B}$, $Hp_{y}^{B}$, $Hp_{x}^{B}$, $Hs^{B}$, $Xs^{B} \}$ and at the $K$  point, the total Hamiltonian with the nearest-neighbor hopping considered in the Slater-Koster formalism reads
\begin{eqnarray}\label{H0-XF}
H_{0}^{ha}=\left(\begin{array}{cc}
h_{0}^{AA} & h_{0}^{AB}\\
{h_{0}^{AB}}^{\dagger} & h_{0}^{BB}
\end{array}\right),
\end{eqnarray}
with
\begin{eqnarray}\label{}
\begin{split}
&h_{0}^{AA}= \\
&\left[\begin{array}{cccccccc}
0 & 0 & 0 & 0 & V_{pp\pi}^{ha} & 0 & 0 & 0\\
0 & 0 & 0 & 0 & 0 & V_{pp\pi}^{ha} & 0 & 0\\
0 & 0 & 0 & V_{pp\sigma}^{ha} & 0 & 0 & -V_{sp\sigma}^{ha} & 0\\
0 & 0 & V_{pp\sigma}^{ha} & \Delta_{p}^{ha} & 0 & 0 & 0 & V_{sp\sigma}^{ha}\\
V_{pp\pi}^{ha} & 0 & 0 & 0 & \Delta_{p}^{ha} & 0 & 0 & 0\\
0 & V_{pp\pi}^{ha} & 0 & 0 & 0 & \Delta_{p}^{ha} & 0 & 0\\
0 & 0 & -V_{sp\sigma}^{ha} & 0 & 0 & 0 & \Delta_{s}^{ha} & V_{ss\sigma}^{ha}\\
0 & 0 & 0 & V_{sp\sigma}^{ha} & 0 & 0 & V_{ss\sigma}^{ha} & \Delta
\end{array}\right],
\end{split}
\nonumber\\
\end{eqnarray}
\begin{eqnarray}\label{}
h_{0}^{AB}=\left[\begin{array}{cccccccc}
-V_{1}^{'} & -iV_{1}^{'} & 0 & 0 & 0 & 0 & 0 & V_{2}^{'}\\
-iV_{1}^{'} & V_{1}^{'} & 0 & 0 & 0 & 0 & 0 & -iV_{2}^{'}\\
0 & 0 & 0 & 0 & 0 & 0 & 0 & 0\\
0 & 0 & 0 & 0 & 0 & 0 & 0 & 0\\
0 & 0 & 0 & 0 & 0 & 0 & 0 & 0\\
0 & 0 & 0 & 0 & 0 & 0 & 0 & 0\\
0 & 0 & 0 & 0 & 0 & 0 & 0 & 0\\
-V_{2}^{'} & iV_{2}^{'} & 0 & 0 & 0 & 0 & 0 & 0
\end{array}\right],
\end{eqnarray}
\begin{eqnarray}\label{}
\begin{split}
& h_{0}^{BB}= \\
&\left[\begin{array}{cccccccc}
0 & 0 & 0 & 0 & V_{pp\pi}^{ha} & 0 & 0 & 0\\
0 & 0 & 0 & 0 & 0 & V_{pp\pi}^{ha} & 0 & 0\\
0 & 0 & 0 & V_{pp\sigma}^{ha} & 0 & 0 & V_{sp\sigma}^{ha} & 0\\
0 & 0 & V_{pp\sigma}^{ha} & \Delta_{p}^{ha} & 0 & 0 & 0 & -V_{sp\sigma}^{ha}\\
V_{pp\pi}^{ha} & 0 & 0 & 0 & \Delta_{p}^{ha} & 0 & 0 & 0\\
0 & V_{pp\pi}^{ha} & 0 & 0 & 0 & \Delta_{p}^{ha} & 0 & 0\\
0 & 0 & V_{sp\sigma}^{ha} & 0 & 0 & 0 & \Delta_{s}^{ha} & V_{ss\sigma}^{ha}\\
0 & 0 & 0 & -V_{sp\sigma}^{ha} & 0 & 0 & V_{ss\sigma}^{ha} & \Delta
\end{array}\right],
\end{split}
\nonumber\\
&&
\end{eqnarray}
where $\Delta_{p}^{ha}$ is the on site energy for the $p$ orbitals of the halogen atom, $\Delta$ ($\Delta_{s}^{ha}$) is the on site energy for the $s$ orbital of textit{X} (halogen) atom, the on site energies for $p$ orbitals of textit{X} atoms are taken to be zero. $V_{pp\pi}^{ha}$ ($V_{pp\sigma}^{ha}$ ) is the hopping between the $p_z$ orbital from textit{X} atom and the $p_z$ orbital from halogen atom in the "shoulder by shoulder" ("head to tail") type. $V_{sp\sigma}^{ha}$ is the hopping between the $p_z$ ($s$) orbital from textit{X} atom and the $s$ ($p_z$) orbital from halogen atom. $V_{ss\sigma}^{ha}$ is the hopping between the $s$ orbital from textit{X} atom and the $s$ orbital from halogen atom. The parameters $V_{1}^{'}$ and $V_{2}^{'}$ take the same expressions as in Sec.II A.

Firstly, we perform the unitary transformation as in Eq.~\eqref{u4x4}, as well as the following unitary transformation
\begin{equation}\label{}
\begin{split} & H\varphi_{1}^{A}=-\frac{1}{\sqrt{2}}\left(Hp_{x}^{A}+iHp_{y}^{A}\right)\\
 & H\varphi_{2}^{B}=\frac{1}{\sqrt{2}}\left(Hp_{x}^{B}-iHp_{y}^{B}\right)\\
 & H\varphi_{3}^{A}=-\frac{1}{\sqrt{2}}\left(Hp_{x}^{A}-iHp_{y}^{A}\right)\\
 & H\varphi_{4}^{B}=-\frac{1}{\sqrt{2}}\left(Hp_{x}^{B}+iHp_{y}^{B}\right)
\end{split}.
\end{equation}
In the new basis $\{  X\varphi_{1}^{A}$, $Xs^{B}$, $H\varphi_{1}^{A}$, $Hs^{B}$, $Xp_{z}^{B}$, $Hp_{z}^{B}$, $X\varphi_{2}^{B}$, $Xs^{A}$, $H\varphi_{2}^{B}$, $Hs^{A}$, $Xp_{z}^{A}$, $Hp_{z}^{A}$,$X\varphi_{3}$, $X\varphi_{4}$, $H\varphi_{3}^{A}$, $H\varphi_{4}^{B} \}$  = $\{ Xp_{y}^{A}$, $Xp_{x}^{A}$, $Xp_{z}^{A}$, $Hp_{z}^{A}$, $Hp_{y}^{A}$, $Hp_{x}^{A}$, $Hs^{A}$, $Xs^{A}$, $Xp_{y}^{B}$, $Xp_{x}^{B}$, $Xp_{z}^{B}$, $Hp_{z}^{B}$, $Hp_{y}^{B}$, $Hp_{x}^{B}$, $Hs^{B}$, $Xs^{B} \}$ $U_{1}^{ha}$, we could rewrite the Hamiltonian in the following block-diagonal form with three decoupled diagonal blocks
\begin{equation}\label{}
H_{1}^{ha}=H_{1,\alpha}^{ha}\oplus H_{1,\beta}^{ha}\oplus H_{1,\gamma}^{ha},
\end{equation}
\begin{eqnarray}\label{}
H_{1,\alpha}^{ha}=\left[\begin{array}{cccccc}
0 & iV_{2} & V_{pp\pi}^{ha} & 0 & 0 & 0\\
-iV_{2} & \Delta & 0 & V_{ss\sigma}^{ha} & 0 & -V_{sp\sigma}^{ha}\\
V_{pp\pi}^{ha} & 0 & \Delta_{p}^{ha} & 0 & 0 & 0\\
0 & V_{ss\sigma}^{ha} & 0 & \Delta_{s}^{ha} & V_{sp\sigma}^{ha} & 0\\
0 & 0 & 0 & V_{sp\sigma}^{ha} & 0 & V_{pp\sigma}^{ha}\\
0 & -V_{sp\sigma}^{ha} & 0 & 0 & V_{pp\sigma}^{ha} & \Delta_{p}^{ha}
\end{array}\right],
\nonumber\\
&&
\end{eqnarray}
\begin{eqnarray}\label{}
H_{1,\beta}^{ha}=\left[\begin{array}{cccccc}
0 & -iV_{2} & V_{pp\pi}^{ha} & 0 & 0 & 0\\
iV_{2} & \Delta & 0 & V_{ss\sigma}^{ha} & 0 & V_{sp\sigma}^{ha}\\
V_{pp\pi}^{ha} & 0 & \Delta_{p}^{ha} & 0 & 0 & 0\\
0 & V_{ss\sigma}^{ha} & 0 & \Delta_{s}^{ha} & -V_{sp\sigma}^{ha} & 0\\
0 & 0 & 0 & -V_{sp\sigma}^{ha} & 0 & V_{pp\sigma}^{ha}\\
0 & V_{sp\sigma}^{ha} & 0 & 0 & V_{pp\sigma}^{ha} & \Delta_{p}^{ha}
\end{array}\right],
\nonumber\\
&&
\end{eqnarray}
\begin{eqnarray}\label{}
H_{1,\gamma}^{ha}=\left[\begin{array}{cccc}
V_{1} & 0 & \frac{V_{pp\pi}^{ha}}{\sqrt{2}} & \frac{V_{pp\pi}^{ha}}{\sqrt{2}}\\
0 & -V_{1} & -\frac{V_{pp\pi}^{ha}}{\sqrt{2}} & \frac{V_{pp\pi}^{ha}}{\sqrt{2}}\\
\frac{V_{pp\pi}^{ha}}{\sqrt{2}} & -\frac{V_{pp\pi}^{ha}}{\sqrt{2}} & \Delta_{p}^{ha} & 0\\
\frac{V_{pp\pi}^{ha}}{\sqrt{2}} & \frac{V_{pp\pi}^{ha}}{\sqrt{2}} & 0 & \Delta_{p}^{ha}
\end{array}\right].
\end{eqnarray}

For the first diagonal block $H_{1,\alpha}^{ha}$, in the presentation $\{X\varphi_{1}^{A},Xs^{B},H\varphi_{1}^{A},Hs^{B},Xp_{z}^{B},Hp_{z}^{B}\}$ its eigenvectors can be written as
\begin{eqnarray}\label{XF-vec1-0}
\begin{split} & |\varepsilon_{i}^{ha}\rangle=\frac{1}{N_{i}^{ha}}\times\\
 & \left[\begin{array}{c}
1\\
\frac{i}{C}\\
\frac{V_{pp\pi}^{ha}}{\varepsilon_{i}^{ha}-\Delta_{p}^{ha}}\\
\frac{i\left[V_{pp\sigma}^{ha}\left(V_{sp\sigma}^{ha2}+V_{pp\sigma}^{ha}V_{ss\sigma}^{ha}\right)-\varepsilon_{i}^{ha}V_{ss\sigma}^{ha}\left(\varepsilon_{i}^{ha}-\Delta_{p}^{ha}\right)\right]}{DC}\\
\frac{-iV_{sp\sigma}^{ha}\left[\Delta_{s}^{ha}V_{pp\sigma}^{ha}-\Delta_{p}^{ha}V_{ss\sigma}^{ha}-\varepsilon_{i}^{ha}\left(V_{pp\sigma}^{ha}-V_{ss\sigma}^{ha}\right)\right]}{DC}\\
\frac{-iV_{sp\sigma}^{ha}\left[V_{sp\sigma}^{ha2}+V_{ss\sigma}^{ha}V_{pp\sigma}^{ha}-\varepsilon_{i}^{ha}\left(\varepsilon_{i}^{ha}-\Delta_{s}^{ha}\right)\right]}{DC}
\end{array}\right],
\end{split}
\end{eqnarray}
with
\begin{equation}\label{XF-vec1-1}
\begin{split}
D\left(\varepsilon_{i}^{ha}\right)\equiv
&\left(\varepsilon_{i}^{ha}-\Delta_{s}^{ha}\right)\left[V_{pp\sigma}^{ha2}-\varepsilon_{i}^{ha}\left(\varepsilon_{i}^{ha}-\Delta_{p}^{ha}\right)\right]+ \\
&\left(\varepsilon_{i}^{ha}-\Delta_{p}^{ha}\right)V_{sp\sigma}^{ha2},
\end{split}
\end{equation}
and
\begin{equation}\label{XF-vec1-2}
C\equiv\frac{V_{2}\left(\varepsilon_{i}^{ha}-\Delta_{p}^{ha}\right)}{V_{pp\pi}^{ha2}-\varepsilon_{i}^{ha}\left(\varepsilon_{i}^{ha}-\Delta_{p}^{ha}\right)}. \end{equation}
Here, $\varepsilon_{i}^{ha}$  and $N_{i}^{ha}\left(i=1,2,\cdots,6\right)$ are the corresponding eigenvalues and the normalization factors, respectively. Therefore, by the unitary transformation
\begin{equation}\label{XF-vec1-t}
\begin{split}
& \left\{ \phi_{1}^{ha},\phi_{2}^{ha},\phi_{3}^{ha},\phi_{4}^{ha},\phi_{5}^{ha},\phi_{6}^{ha}\right\} \\
&=\left\{ X\varphi_{1}^{A},Xs^{B},H\varphi_{1}^{A},Hs^{B},Xp_{z}^{B},Hp_{z}^{B}\right\} U_{\alpha},
\end{split}
 \end{equation}
with $U_{\alpha}=\{ |\varepsilon_{i}^{ha}\rangle\} _{i=1,2,\cdots,6}\equiv\{ u_{ji}^{\alpha}\}$, the above $6\times6$ block $H_{1,\alpha}^{ha}$ is diagonalized.

From our FP calculations [Fig.~\ref{fig:Bands}(b)], the main components of the band around the Dirac points and the Fermi level come from the $Xp_{x}$  and $Xp_{y}$ orbitals, mixed with a small amount of the $Hp_{x}$ and $Hp_{y}$ orbitals as well as $Xs$ orbital. The orbital features are identical with the eigenvectors of $\varepsilon_{1}^{ha}$. When we take its eigenvalue as the Fermi energy $E_F$. Following similar procedures as in the previous section, we can obtain the eigenvalues up to the second-order correction and the eigenvectors up to the first-order correction with
\begin{eqnarray}\label{HX-eigen-complex}
\varepsilon_{1}^{ha}=\frac{1}{2}\left(\Delta^{'}+\sqrt{\Delta^{'2}+4V_{2}^{2}-2\frac{\Delta^{'}V_{pp\pi}^{ha2}}{\varepsilon-\Delta_{p}^{ha}}+\frac{V_{pp\pi}^{ha4}}{\left(\varepsilon-\Delta_{p}^{ha}\right)^{2}}}\right), \nonumber\\
&&
\end{eqnarray}
where
\begin{eqnarray}\label{XF-egienvalue-1}
\begin{split}
& \Delta^{'}=\Delta-\\
&\frac{\varepsilon_{1}^{ha^{0}2}\left(V_{ss\sigma}^{ha2}+V_{sp\sigma}^{ha2}\right)-\varepsilon_{1}^{ha^{0}}\left(\Delta_{p}^{ha}V_{ss\sigma}^{ha2}+\Delta_{s}^{ha}V_{sp\sigma}^{ha2}\right)}{D\left(\varepsilon_{i}^{ha^{0}}\right)}+\\
&+\frac{\left(V_{sp\sigma}^{ha2}+V_{ss\sigma}^{ha}V_{pp\sigma}^{ha}\right)}{D\left(\varepsilon_{i}^{ha^{0}}\right)}, \end{split}
\nonumber\\
&&
\end{eqnarray}
\begin{eqnarray}\label{HX-eigen1}
\varepsilon_{1}^{ha^{0}}=\frac{1}{2}\left(\Delta+\sqrt{\Delta^{2}+4V_{2}^{2}-2\frac{\Delta V_{pp\pi}^{ha2}}{\varepsilon-\Delta_{p}^{ha}}+\frac{V_{pp\pi}^{ha4}}{\left(\varepsilon-\Delta_{p}^{ha}\right)^{2}}}\right), \nonumber\\
&&
\end{eqnarray}
and
\begin{eqnarray}\label{HX-eigen0}
\varepsilon=\frac{1}{2}\left(\Delta+\sqrt{\Delta^{2}+4V_{2}^{2}}\right).
\end{eqnarray}
Up to this point, we have found the low-energy eigenvalue $\varepsilon_{1}^{ha}$ and the corresponding basis $\phi_{1}^{ha}$.
Again, in order to capture the essential physics, we simply the above expressions by taking only the zeroth-order terms. So in the following, we take $\varepsilon_{1}^{ha}=\varepsilon_{1}^{ha^{0}}$  and omit the correction with $\{ Hs^{B},Xp_{z}^{B},Hp_{z}^{B}\}$  for the eigenvector $\{ |\varepsilon_{1}^{ha}\rangle\}$. Consequently, the eigenvector has the following form in the basis $\{ X\varphi_{1}^{A},Xs^{B},H\varphi_{1}^{A}\}$
\begin{eqnarray}\label{XF-vec1}
|\varepsilon_{1}^{ha}\rangle=\frac{1}{n_{1}^{ha}}\left[\begin{array}{c}
1\\
-i\frac{V_{2}}{\varepsilon_{1}^{ha^{0}}-\Delta}\\
\frac{V_{pp\pi}^{ha}}{\varepsilon_{1}^{ha^{0}}-\Delta_{p}^{ha}}
\end{array}\right]\equiv\left[\begin{array}{c}
u_{11}^{ha}\\
u_{21}^{ha}\\
u_{31}^{ha}
\end{array}\right],
\end{eqnarray}
with $n_{1}^{ha}$ being a normalization constant, and the eigenvalue $\varepsilon_{1}^{ha^{0}}$ is given in Eqs.~(\ref{HX-eigen1}) and (\ref{HX-eigen0}).

The eigenvalues of the second diagonal block $H_{1,\beta}^{ha}$  are denoted as $\varepsilon_{6+i}^{ha}$ $(i=1,2,\cdots,6)$, and one finds that $\varepsilon_{6+i}^{ha}=\varepsilon_{i}^{ha}$ $(i=1,2,\cdots,6)$, where $\varepsilon_{i}^{ha}$  are eigenvalues of $H_{1,\alpha}^{ha}$. Through similar procedures, the low-energy eigenvector $\{|\varepsilon_{7}^{ha}\rangle\}$  has the following simple form in the basis $\{ X\varphi_{2}^{B},Xs^{A},H\varphi_{2}^{B}\}$:
\begin{eqnarray}\label{XF-vec7}
|\varepsilon_{7}^{ha}\rangle=\frac{1}{n_{1}^{ha}}\left[\begin{array}{c}
1\\
i\frac{V_{2}}{\varepsilon_{1}^{ha^{0}}-\Delta}\\
\frac{V_{pp\pi}^{ha}}{\varepsilon_{1}^{ha^{0}}-\Delta_{p}^{ha}}
\end{array}\right]=\left[\begin{array}{c}
u_{11}^{ha}\\
-u_{21}^{ha}\\
u_{31}^{ha}
\end{array}\right].
\end{eqnarray}
The third diagonal block $H_{1,\gamma}^{ha}$ are of high energy hence is not of interest here.

From the above analysis, the low-energy states $\phi_{1}^{ha}$  and $\phi_{7}^{ha}$ constitute the low-energy Hilbert subspace. They have the following explicit forms:
\begin{equation}\label{simple-Hilbert-XF-K}
\begin{split}
\phi_{1}^{ha}=
& u_{11}^{ha}\left[-\frac{1}{\sqrt{2}}\left(Xp_{x}^{A}+iXp_{y}^{A}\right)\right]+u_{21}^{ha}Xs^{B} \\
& +u_{31}^{ha}\left[-\frac{1}{\sqrt{2}}\left(Hp_{x}^{A}+iHp_{y}^{A}\right)\right], \\
\phi_{7}^{ha}=
& u_{11}^{ha}\left[\frac{1}{\sqrt{2}}\left(Xp_{x}^{B}-iXp_{y}^{B}\right)\right]-u_{21}^{ha}Xs^{A}  \\ & +u_{31}^{ha}\left[\frac{1}{\sqrt{2}}\left(Hp_{x}^{B}-iHp_{y}^{B}\right)\right].
\end{split}
\end{equation}
Again we perform the small $\vec{k}$  expansion in the above low-energy Hilbert subspace around $K$ point by $\vec{k}\rightarrow\vec{k}+K$  and keep the first-order terms in $\vec{k}$,
\begin{eqnarray}\label{XF-HK}
H_{K}=\left(\begin{array}{cc}
0 & v_{F}k_{-}\\
v_{F}k_{+} & 0
\end{array}\right),
\end{eqnarray}
with $v_{F}$  the Fermi velocity
\begin{equation}\label{vf-XF}
v_{F}=\frac{\sqrt{3}a}{2}\left[\frac{1}{2}\mid u_{11}^{ha}\mid^{2}\left(V_{pp\sigma}-V_{pp\pi}\right)+\mid u_{21}^{ha}\mid^{2}V_{ss\sigma}\right].
\end{equation}
Note that for the textit{X}-halide systems, $\mid u_{11}^{ha}\mid^{2}$  is much larger than $\mid u_{21}^{ha}\mid^{2}$  and $\mid u_{31}^{ha}\mid^{2}$.
Either following similar procedures, or via the inversion symmetry (or time-reversal symmetry ), one can obtain the low-energy Hilbert subspace and and the low-energy effective Hamiltonian around the $K'$ point. Finally the basis for low-energy Hilbert subspace can be summarized as
\begin{equation}\label{simple-Hilbert-XF}
\begin{split}
\phi_{1}^{ha}=
& u_{11}^{ha}\left[-\frac{1}{\sqrt{2}}\left(Xp_{x}^{A}+i\tau_{z}Xp_{y}^{A}\right)\right]+u_{21}^{ha}\tau_{z}Xs^{B} \\
& +u_{31}^{ha}\left[-\frac{1}{\sqrt{2}}\left(Hp_{x}^{A}+i\tau_{z}Hp_{y}^{A}\right)\right],  \\
\phi_{7}^{ha}=
&u_{11}^{ha}\left[\frac{1}{\sqrt{2}}\left(Xp_{x}^{B}-i\tau_{z}Xp_{y}^{B}\right)\right]-u_{21}^{ha}\tau_{z}Xs^{A}\\
&+u_{31}^{ha}\left[\frac{1}{\sqrt{2}}\left(Hp_{x}^{B}-i\tau_{z}Hp_{y}^{B}\right)\right].\\
\end{split}
\end{equation}
and the low-energy effective Hamiltonian without SOC reads
\begin{equation}\label{Heff-XF}
H_{\tau}=v_{F}\left(k_{x}\sigma_{x}+\tau_{z}k_{y}\sigma_{y}\right),
\end{equation}
where Pauli matrices $\sigma$ denote the orbital basis degree of freedom, and $\tau_{z}$ labels the two valleys $K$ and $K'$. Note that under the space reversal operation $P=\sigma_{x}\tau_{x}$ and the time-reversal operation $T=\tau_{x}\hat{K}$, the above low-energy effective Hamiltonian Eq.~\eqref{Heff-XF} is also invariant.

\subsection{Low-energy effective Hamiltonian involving SOC}

In a similar way as in Sec. II B, we obtain an on-site SOC in the spinful low-energy Hilbert subspace $\{ \phi_{1},\phi_{7}\}\otimes\{ \uparrow,\downarrow\}$,
\begin{equation}\label{soc-XF}
H_{so}=\lambda_{so}\tau_{z}\sigma_{z}s_{z},
\end{equation}
\begin{equation}\label{lamdaso-XF}
\lambda_{so}=\frac{1}{2}\mid u_{11}^{ha}\mid^{2}\xi_{0}^{X}+\frac{1}{2}\mid u_{31}^{ha}\mid^{2}\xi_{0}^{ha},
\end{equation}
where $u_{11}^{ha}$  and $u_{31}^{ha}$  are given in Eq.~\eqref{XF-vec1}, and $\xi_{0}^{X}$ ($\xi_{0}^{ha}$) is the magnitude of atomic SOC of pnictogen (halogen). It should be noted that due to the presence of major $p_x$ and $p_y$ orbital components, the first-order on-site effective SOC also dominates in the textit{X}-halide systems. Equation~\eqref{XF-vec1} explains the tendency that the $\lambda_{so}$ increases with the atomic number of halogen for the same pnictogen element, as shown in Table ~\ref{tab:XF-parameter}.

From Eqs.~(\ref{Heff-XF}) and (\ref{soc-XF}), we obtain the generic low-energy effective Hamiltonian around the Dirac points acting on the low-energy Hilbert subspace $\{\phi_{1},\phi_{7}\}\otimes\{\uparrow,\downarrow\}$
\begin{equation}\label{Heffsoc-XF}
H_\text{eff}=H_{\tau}+H_{so}=v_{F}\left(k_{x}\sigma_{x}+\tau_{z}k_{y}\sigma_{y}\right)+\lambda_{so}\tau_{z}\sigma_{z}s_{z},
\end{equation}
where Fermi velocity $v_{F}$ and magnitude of intrinsic effective SOC $\lambda_{so}$ are given in Eqs.~(\ref{vf-XF}) and (\ref{lamdaso-XF}), and their values are listed in Table ~\ref{tab:XF-parameter}. One notes that this Hamiltonian is also invariant under both the space-inversion symmetry and time-reversal symmetry with $T=is_y\tau_{x}\hat{K}$.

The two model parameters $v_F$ and $\lambda_{so}$ for halides obtained by fitting the band dispersions of the FP results are listed in Table \ref{tab:XF-parameter}.

\begin{table*}
\caption{Values of two model parameters $v_{F}$ and $\lambda_{so}$ for honeycomb textit{X}-halide (\textit{X} = N-Bi) monolayers obtained from FP calculations. Note that $\lambda_{so}=E_{g}/2$, with $E_{g}$ the gap opened by SOC at the Dirac point.}\label{tab:XF-parameter}
\begin{ruledtabular}
\begin{tabular}{cccccccccccccc}
& system & $v_{F}\left(10^{5}m/s\right)$ & $\lambda_{so}\left(eV\right)$ & system & $v_{F}\left(10^{5}m/s\right)$ & $\lambda_{so}\left(eV\right)$\\
\hline
& NF & 5.5 & $8.5\times10^{-3}$ & NBr & 4.2 & $19\times10^{-3}$\\
& PF & 7.2 & $13\times10^{-3}$  & PBr & 8.0 & $17\times10^{-3}$ \\
& AsF &7.3 & $80\times10^{-3}$  & AsBr & 8.2 & $98\times10^{-3}$\\
& SbF &6.6 & 0.16 & SbBr & 7.7 & 0.20\\
& BiF &7.2 & 0.55 & BiBr & 7.3  &  0.65 \\
& NCl & 4.3 & $9.7\times10^{-3}$ & NI & 3.8 & $28\times10^{-3}$\\
& PCl & 7.8 & $17\times10^{-3}$  & PI & 8.1 & $19\times10^{-3}$\\
& AsCl &8.0 & $95\times10^{-3}$  & AsI & 9.1 & $0.10$\\
& SbCl &7.3 & 0.19 & SbI &7.7 & 0.21\\
& BiCl &6.9 & 0.56 & BiI &7.7 & 0.65\\
\end{tabular}
\end{ruledtabular}
\end{table*}

\section{A simple spinful lattice Hamiltonian for the honeycomb textit{X}-hydride/halide (\textit{X} = N-Bi) monolayers family}

For the purpose of studying the topological properties of the honeycomb textit{X}-hydride/halide (\textit{X} = N-Bi) monolayers family, as well as their edge states, it is convenient to work with a lattice Hamiltonian via lattice regularization of the low-energy continuum models (Eq.~\eqref{Heffsoc} and Eq.~\eqref{Heffsoc-XF}). Taking into account the main physics involving $p_x$ and $p_y$ orbitals, we construct the following spinful lattice Hamiltonian for the 2D honeycomb textit{X}-hydride/halide (\textit{X} = N-Bi) monolayers
\begin{equation}\label{Hamiltonian-lattice-eff}
\begin{split}
H=
& \sum_{\langle i,j\rangle;\alpha,\beta=p_{x},p_{y}}t_{ij}^{\alpha\beta}c_{i\alpha}^{\dagger}c_{j\beta} \\
&+\sum_{i;\alpha,\beta=p_{x},p_{y};\sigma,\sigma^{'}=\uparrow,\downarrow}\lambda_{\sigma,\sigma^{'}}^{\alpha\beta} c_{i\alpha\sigma}^{\dagger}c_{i\beta\sigma^{'}}s_{\sigma,\sigma^{'}}^{z}, \end{split}
\end{equation}
where $\langle i,j\rangle$  means $i$ and $j$ sites are nearest neighbors, $\alpha$ and $\beta$ are the orbital indices. The first term is the hopping term and the second one is the on-site SOC term.

After Fourier transformation of the above lattice Hamiltonian, its energy spectrum over the entire Brillouin zone can be obtained. Since here spin is good quantum number, we can divide the model Hamiltonian into two sectors for spin up and spin down separately. For each sector, the corresponding model Hamiltonian reads
\begin{eqnarray}\label{model_Hamiltonian_up}
H^{\uparrow}\left(k\right)=\left[\begin{array}{cccc}
0 & -\frac{i\xi_{0}}{2} & h_{xx}^{AB}\left(k\right) & h_{xy}^{AB}\left(k\right) \\
       &0  & h_{xy}^{AB}\left(k\right) &h_{yy}^{AB}\left(k\right) \\
       &        & 0 & -\frac{i\xi_{0}}{2} \\
\multicolumn{2}{c}{\raisebox{1.3ex}[0pt]{$\dagger$}}
                &        &0
\end{array}\right],
\end{eqnarray}
\begin{eqnarray}\label{model_Hamiltonian_dn}
H^{\downarrow}\left(k\right)=\left[\begin{array}{cccc}
0 & \frac{i\xi_{0}}{2} & h_{xx}^{AB}\left(k\right) & h_{xy}^{AB}\left(k\right) \\
       &0  & h_{xy}^{AB}\left(k\right) &h_{yy}^{AB}\left(k\right) \\
       &        & 0 & \frac{i\xi_{0}}{2} \\
\multicolumn{2}{c}{\raisebox{1.3ex}[0pt]{$\dagger$}}
                &        &0
\end{array}\right],
\end{eqnarray}
where
\begin{equation*}
\begin{split}
h_{xx}^{AB}\left(k\right)\equiv
& \frac{1}{2}\left(3V_{pp\sigma}+V_{pp\pi}\right)\cos\left(\frac{k_{x}}{2}\right)\exp\left(i\frac{k_{y}}{2\sqrt{3}}\right) \\
& +V_{pp\pi}\exp\left(-i\frac{k_{y}}{\sqrt{3}}\right), \\
\end{split}
\end{equation*}
\begin{equation*}
h_{xy}^{AB}\left(k\right)\equiv i\frac{\sqrt{3}}{2}\left(V_{pp\sigma}-V_{pp\pi}\right)\sin\left(\frac{k_{x}}{2}\right)\exp\left(i\frac{k_{y}}{2\sqrt{3}}\right),
\end{equation*}
and
\begin{equation*}
\begin{split}
h_{yy}^{AB}\left(k\right)\equiv
& \frac{1}{2}\left(V_{pp\sigma}+3V_{pp\pi}\right)\cos\left(\frac{k_{x}}{2}\right)\exp\left(i\frac{k_{y}}{2\sqrt{3}}\right) \\
& +V_{pp\sigma}\exp\left(-i\frac{k_{y}}{\sqrt{3}}\right). \\
\end{split}
\end{equation*}
For simplicity, we choose the lattice constant $a=1$. The on-site energies for $p$ orbitals are taken to be zero. Near the $K$ and $K'$ points, the above model Hamiltonian reduces to the low-energy effective Hamiltonian [Eq.~\eqref{Heffsoc} and ~\eqref{Heffsoc-XF}] with $v_{F}=\frac{\sqrt{3}a}{4}\left(V_{pp\sigma}-V_{pp\pi}\right)$ and $\lambda_{so}=\xi_{0}/2$.

\begin{figure}
\includegraphics[width=3.5in]{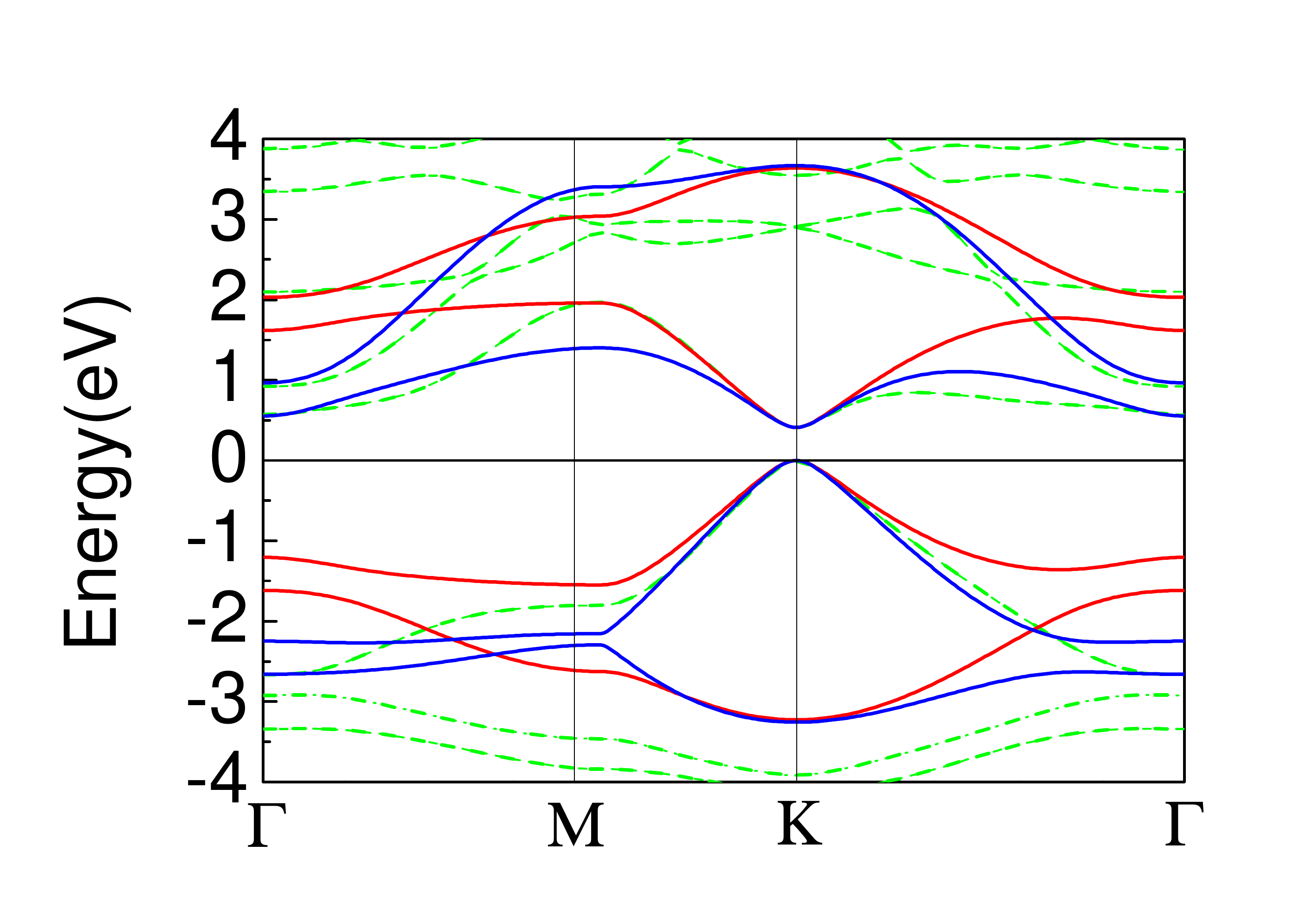}
\caption{(Color online). A comparison of the band structures for monolayer SbH calculated using FP and TB methods with SOC . The dashed green curve is the FP result.
The solid red and blue curves are the TB model results. The red curve is with the NN hopping only, while the blue curve also includes the NNN hopping terms. For the NN case, the parameters are taken as $V_{pp\sigma}=1.68 eV$, $V_{pp\pi}=-0.60 eV$. For the NNN case, the parameters are taken as $V_{pp\sigma}=1.69 eV$, $V_{pp\pi}=-0.62 eV$, $V_{pp\sigma}^{NNN}=0 eV$, $V_{pp\pi}^{NNN}=-0.23 eV$. For both cases, $\lambda_{so}=0.21 eV$. The superscript NNN means the next-nearest-neighbor hoping. The Fermi level is set to zero.}\label{fig:TBvsFP}
\end{figure}

Taking SbH as an example, we compare the results from FP calculations and from the lattice models. As shown in Fig.~\ref{fig:TBvsFP}, there is a good agreement between the two results around the $K$ point. The fitting away from $K$ point can be improved by including hopping terms between far neighbors. In Fig.~\ref{fig:TBvsFP}, we also show the result with NNN hopping, for which a fairly good agreement with the FP low-energy bands over the whole Brillouin zone can be achieved.

\section{Discussion and Summary}

We have obtained the low-energy effective Hamiltonian for the textit{X}-hydride and textit{X}-halide (\textit{X} = N-Bi) family of materials, which is analogous to the Kane-Mele model proposed for the QSH effect in graphene.~\cite{kane2005a} The important difference is that in Kane-Mele model the effective SOC is of second-order NNN type, which is much weaker than the on-site SOC in our systems. The SOC term in our Hamiltonian opens a large nontrivial gap at the Dirac points. From $K$ to $K'$ the mass term changes sign for each spin species and the band is inverted. As a result, the QSH effect can be realized in the textit{X}-hydride and textit{X}-halide (\textit{X} = N-Bi) monolayers. Some of these materials, such as BiH/BiF, have record huge SOC gap with magnitude around 1 eV, far higher than the room-temperature energy scale, hence making their detection much easier.

On the experimental side, the buckling honeycomb Bi(111) monolayer and film have been manufactured via molecular beam epitaxy (MBE).~\cite{Hirahara2011,Yang2012,Sabater2013} On the other hand, chemical functionalization of 2D materials is a powerful tool to create new materials with desirable features, such as modifying graphene into graphane, graphone, and fluorinated graphene via H and F, respectively.~\cite{graphane} Therefore, it is very promising that Bi-Hydride/Halide monolayer, the huge gap QSH insulators, may be synthesized by chemical reaction in the solvents or by the exposure of the Bi (111) monolayer and film to the atomic or molecular gases. It is noted that even though one side (full passivation) instead of both sides (alternating passivation) of Bi(111) bilayers is passivated, the band structure is almost unchanged and the topology properties remain nontrivial. This will provide more freedom to realize these kinds of materials.

It is known that the low-energy Hilbert space for graphene consists of the $p_{z}$ orbital from carbon atoms. In that system, the SOC term from NNN second-order processes is vanishingly small, and the on-site SOC as well as the nearest neighbor SOC are forbidden by symmetry constraint.
In contrast, for the honeycomb textit{X}-hydride/halide monolayers, $p_{x}$ and $p_{y}$ orbitals from the group V elements constitute the low-energy Hilbert subspace. In fact, this represents the first class of materials for which the Dirac fermion physics is associated with $p_{x}$ and $p_{y}$ orbitals. Because of this, the effective on-site SOC can has nonzero matrix elements and results in the huge SOC gap at the Dirac points.

\begin{figure}
\includegraphics[width=3.5in]{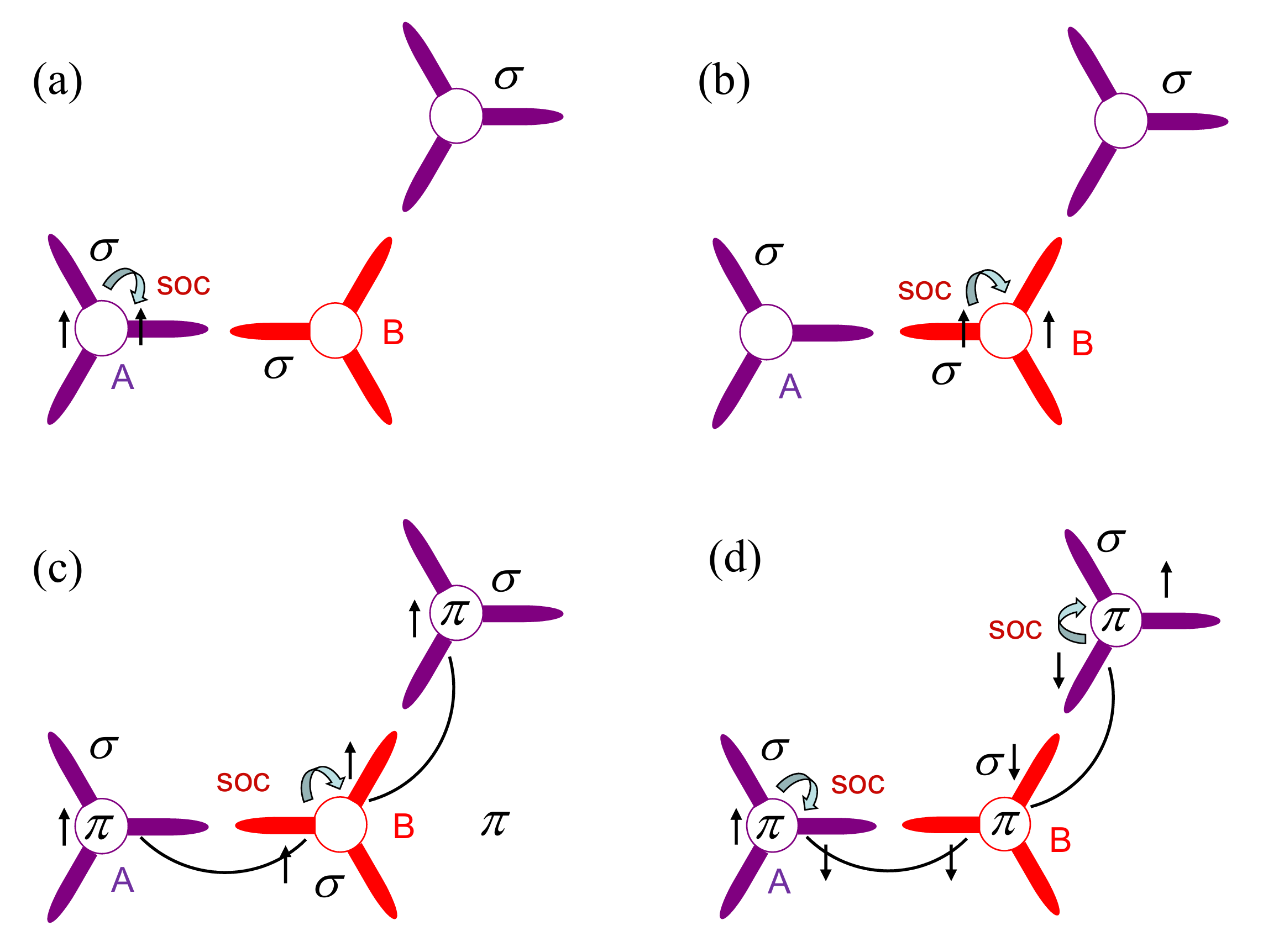}
\caption{(Color online). The leading-order effective SOC processes in textit{X}-hydride or textit{X}-halide (\textit{X} = N-Bi), silicene and graphene. (a) and (b) Sketches of the huge effective on-site SOC in textit{X}-hydride systems and textit{X}-halide systems. (c) Sketch of the effective SOC from NNN hopping processes caused by
the buckling in silicene.  (d) Sketch of the second-order effective SOC from NNN hopping processes in graphene.}\label{fig:soc_hopping}
\end{figure}

The leading-order effective SOC processes in the textit{X}-hydride and textit{X}-halide systems, silicene, and graphene are schematically shown in Fig.~\ref{fig:soc_hopping}. As shown in Figs.~\ref{fig:soc_hopping}4(a) and 4(b), the representative leading-order effective SOC processes around the $K$ point in the honeycomb textit{X}-hydride and textit{X}-halide monolayers are
\begin{equation}\label{soc_hop}
\begin{split}
& |p_{+\uparrow}^{A}\rangle\overset{\lambda_{so}}{\longrightarrow}|p_{+\uparrow}^{A}\rangle,\qquad|p_{+\downarrow}^{A}\rangle\overset{-\lambda_{so}}{\longrightarrow}|p_{+\downarrow}^{A}\rangle, \\
& |p_{-\uparrow}^{B}\rangle\overset{-\lambda_{so}}{\longrightarrow}|p_{-\uparrow}^{B}\rangle,\qquad|p_{-\downarrow}^{B}\rangle\overset{\lambda_{so}}{\longrightarrow}|p_{-\downarrow}^{B}\rangle, \end{split}
\end{equation}
where $\lambda_{so}$ represents the atomic spin-orbit interaction strength, which is given in Eq.~\eqref{lamdaso} for textit{X}-hydride systems and Eq.~\eqref{lamdaso-XF} for textit{X}-halide systems. In a Hilbert subspace consisting of $p_x$ and $p_y$ orbitals, such effective SOC arises in the first-order on-site processes, which leads to its huge magnitude.

As for silicene, which has a low-buckled structure, the typical leading-order SOC is from the (first-order) NNN processes,~\cite{liu_low-energy_2011} as shown in Fig.~\ref{fig:soc_hopping}(c),
\begin{equation}\label{1st}
\begin{split}
&|p_{z\uparrow}^{A}\rangle\overset{V}{\longrightarrow}|p_{-\uparrow}^{B}\rangle\overset{-\frac{\xi_{0}}{2}}{\longrightarrow}|p_{-\uparrow}^{B}\rangle\overset{V}{\longrightarrow}|p_{z\uparrow}^{A}\rangle, \\
&|p_{z\downarrow}^{A}\rangle\overset{V}{\longrightarrow}|p_{-\downarrow}^{B}\rangle\overset{\frac{\xi_{0}}{2}}{\longrightarrow}|p_{-\downarrow}^{B}\rangle\overset{V}{\longrightarrow}|p_{z\downarrow}^{A}\rangle, \\
&|p_{z\uparrow}^{B}\rangle\overset{V}{\longrightarrow}|p_{+\uparrow}^{A}\rangle\overset{\frac{\xi_{0}}{2}}{\longrightarrow}|p_{+\uparrow}^{A}\rangle\overset{V}{\longrightarrow}|p_{z\uparrow}^{B}\rangle,
\\
&|p_{z\downarrow}^{B}\rangle\overset{V}{\longrightarrow}|p_{+\downarrow}^{A}\rangle\overset{-\frac{\xi_{0}}{2}}{\longrightarrow}|p_{+\downarrow}^{A}\rangle\overset{V}{\longrightarrow}|p_{z\downarrow}^{B}\rangle,
\end{split}
\end{equation}
where $V$ is the nearest-neighbor direct hopping amplitude and $\xi_{0}$ represents the atomic intrinsic SOC strength. The whole process can be divided into three steps. For example, we consider the $p_z^A$ orbital. Firstly, due to the low-buckled structure, $p_z^{A}$ couples to $p_{-}^{B}$. Carriers in $p_z^{A}$ orbital then hop to the nearest neighbor $p_{-}^B$ orbital. Secondly, the atomic intrinsic SOC shifts the energy of the spin up and spin down carriers by $\mp\frac{\xi_0}{2}$. In the third step, carriers in the $p_{-}^{B}$ orbital hop to another nearest-neighbor $p_z^A$ orbital, making the resulting effective SOC an NNN process and of first order in $\xi_0$.

As for graphene, around Dirac point, the leading-order effective SOC is from (second-order) NNN effective SOC process, as shown in Fig.~\ref{fig:soc_hopping}(d):
\begin{equation}
\begin{split}
&|p_{z\uparrow}^{A}\rangle\overset{\xi_{0}/\sqrt{2}}{\longrightarrow}|p_{+\downarrow}^{A}\rangle\overset{V}{\longrightarrow}|s_{\downarrow}^{B}\rangle\overset{V}{\longrightarrow}|p_{+\downarrow}^{A}\rangle\overset{\xi_{0}/\sqrt{2}}{\longrightarrow}|p_{z\uparrow}^{A}\rangle,
\\
&|p_{z\downarrow}^{B}\rangle\overset{\xi_{0}/\sqrt{2}}{\longrightarrow}|p_{-\uparrow}^{B}\rangle\overset{V}{\longrightarrow}|s_{\uparrow}^{A}\rangle\overset{V}{\longrightarrow}|p_{-\uparrow}^{B}\rangle\overset{\xi_{0}/\sqrt{2}}{\longrightarrow}|p_{z\downarrow}^{B}\rangle.
\end{split}
\end{equation}
During the whole NNN hopping process, the atomic SOC appears twice, making the effective SOC second order in $\xi_0$ and hence much weaker.

In summary, using the TB method and the FP calculation, we have derived the low-energy effective Hilbert subspace and Hamiltonian for the honeycomb textit{X}-hydride/halide monolayers materials. These 2D group-V honeycomb lattice materials have the same low-energy effective Hamiltonian due to their same $D_{3d}$ point group symmetry and the same $D_{3}$ small group at the $K$ and $K'$ points. The low-energy model contains two key parameters $v_{F}$ and $\lambda_{so}$. We have obtained their analytic expressions and also their numerical values by fitting the FP calculations. Moreover, we have found that the low-energy Hilbert subspace consists of $p_{x}$ and $p_{y}$ orbitals from the group-V elements, which is a key reason for the  huge SOC gap. This feature is distinct from the group-IV honeycomb lattice monolayers such as silicene and graphene. Finally, we construct a spinful lattice Hamiltonian for these materials. Our results will be useful for further investigations of this intriguing class of materials.

\begin{acknowledgments}
This work was supported by the MOST Project of China (Nos. 2014CB920903, 2010CB833104, and 2011CBA00100), the National Natural Science Foundation of China (Grant Nos. 11225418, 51171001, and 11174337), SUTD-SRG-EPD2013062, and the Specialized Research Fund for the Doctoral Program of Higher Education of China (Grant No. 20121101110046). Cheng-Cheng Liu was supported Excellent young scholars Research Fund of Beijing Institute of Technology (Grant No. 2014CX04028).
\end{acknowledgments}

{\sl Note added}  Recently, we notice another relevant work~\cite{Zhang2014} discussing effective models of a honeycomb lattice with $p_x$ and $p_y$ orbitals.

%\bibliography{TB}

\begin{references}

\bibitem{Gelm2007} A. K. Gelm and K. S. Novoselov, Nat. Mater. {\bf 6}, 183-191 (2007).

\bibitem{RMP.Neto.2009} A. H. Castro Neto, F. Guinea, N. M. R. Peres, K. S. Novoselov, and A. K. Geim, Rev. Mod. Phys. 81, 109 (2009).

\bibitem{vogt_silicene:_2012} P. Vogt, P. De Padova, C. Quaresima, J. Avila, E. Frantzeskakis, M. C. Asensio, A. Resta, B. Ealet, and G. Le Lay, Phys. Rev. Lett. \textbf{108}, 155501 (2012).
\bibitem{chen_evidence_2012} L. Chen, C.-C. Liu, B. Feng, X. He, P. Cheng, Z. Ding, S. Meng, Y. Yao, and K. Wu, Phys. Rev. Lett. \textbf{109}, 056804 (2012).

%\bibitem{Bianco_Stability_2013} E. Bianco, S. Butler, S. Jiang, O. D. Restrepo, W. %Windl, and J. E. Goldberger, ACS Nano 7, 4414 (2013).


\bibitem{Song2014} Z. Song, C.-C. Liu, J. Yang, J. Han, B. Fu, Y. Yang, Q. Niu, J. Lu, and Y.G. Yao, arXiv:cond-mat/1402.2399.

\bibitem{Wu2007} C. Wu, D. Bergman, L. Balents, and S. Das Sarma, Phys. Rev. Lett. {\bf 99}, 070401 (2007).

\bibitem{Wu2008} C. Wu, Phys. Rev. Lett. {\bf 100}, 200406 (2008).


%\bibitem{Zhang2010} S. Zhang, H.-h. Hung, and C. Wu, Phys. Rev. A {\bf 82}, 053618 (2010)

\bibitem{Hasan_2010} M. Z. Hasan and C. L. Kane, Rev. Mod. Phys. 82, 3045 (2010).

\bibitem{Qi_2010} X. Qi and S. Zhang, Rev. Mod. Phys. 83, 1057 (2011).

\bibitem{Yan_2012} B. Yan and S.-C. Zhang, Rep. Prog. Phys. 75, 096501 (2012).

\bibitem{kane2005a} C. L. Kane and E. J. Mele, Phys. Rev. Lett. {\bf 95}, 226801 (2005).

\bibitem{kane2005b} C. L. Kane and E. J. Mele, Phys. Rev. Lett. {\bf 95}, 146802 (2005).

\bibitem{yao2007} Y. G. Yao, F. Ye, X. L. Qi, S. C. Zhang, and Z. Fang, Phys. Rev. B {\bf 75}, 041401(R) (2007).

\bibitem{PhysRevB.Min.2006} H. Min, J. E. Hill, N. A. Sinitsyn, B. R. Sahu, L. Kleinman, and A. H. MacDonald, Phys. Rev. B {\bf 74}, 165310 (2006).

\bibitem{Konschuh2010} S. Konschuh, M. Gmitra, and J. Fabian, Phys. Rev. B {\bf 82}, 245412 (2010).

\bibitem{Science.Bernevig.2006} B. A. Bernevig, T. L. Hughes, and S. C. Zhang, Science {\bf 314}, 1757-1761 (2006).

\bibitem{Science.318.766} M. K$\ddot{o}$nig, S. Wiedmann, C. Br$\ddot{u}$ne, A. Roth, H. Buhmann, L. W. Molenkamp, X. L. Qi, and S.C. Zhang, Science {\bf 318}, 766-770 (2007).

\bibitem{PhysRevLett.Liu.2008} C. X. Liu, T. L. Hughes, X. L. Qi, K. Wang, and S. C. Zhang, Phys. Rev. Lett. {\bf 100}, 236601 (2008).

\bibitem{PRL.Knez.2011} I. Knez, R.R. Du, and G. Sullivan, Phys. Rev. Lett. 107, 136603 (2011).

\bibitem{PRL.Knez.2012} I. Knez, R.R. Du, and G. Sullivan, Phys. Rev. Lett. 109, 186603 (2012).

\bibitem{PhysRevLett.Murakami.2006} S. Murakami, Phys. Rev. Lett. {\bf 97}, 236805 (2006).

\bibitem{Liu2011} Z. Liu, C. X. Liu, Y.S. Wu, W.H. Duan, F. Liu, and Jian Wu, Phys. Rev. Lett. {\bf 107}, 136805 (2011).

\bibitem{Hirahara2011}  T. Hirahara, G. Bihlmayer, Y. Sakamoto, M. Yamada, H. Miyazaki, S.I. Kimura, S. Bl$\ddot{u}$gel, and S. Hasegawa, Phys. Rev. Lett. {\bf 107}, 166801 (2011).

\bibitem{liu_quantum_2011} C.-C. Liu, W. Feng, and Y. Yao, Phys. Rev. Lett. {\bf 107}, 076802 (2011).

\bibitem{liu_low-energy_2011} C.-C. Liu, H. Jiang, and Y. Yao, Phys. Rev. B {\bf 84}, 195430 (2011).

\bibitem{PRX.Weeks.2011} C. Weeks, J. Hu, J. Alicea, M. Franz, and R. Wu, Phys. Rev. X 1, 021001 (2011).

\bibitem{Xu2013} Y. Xu, B. Yan, H.-J. Zhang, J. Wang, G. Xu, P. Tang, W. Duan, and S.-C. Zhang, Phys. Rev. Lett. {\bf 111}, 136804 (2013).

\bibitem{Wang2013} Z.F. Wang, Z. Liu, and F. Liu, Nat. Commun. {\bf 4}, 1471 (2013).

\bibitem{Yang2012} F. Yang,L. Miao, Z. F. Wang, M.-Y. Yao, F. Zhu, Y. R. Song, M.-X. Wang, J.-P. Xu, A. V. Fedorov, Z. Sun, G. B. Zhang, C. Liu, F. Liu, D. Qian, C. L. Gao, and J.-F. Jia, Phys. Rev. Lett. {\bf 109}, 016801 (2012).

\bibitem{Zhou2014} M. Zhou, W. Ming, Z. Liu, Z. Wang, Y.G. Yao, and F. Liu, arXiv:cond-mat/1401.3392.

\bibitem{Slater1954} J. C. Slater and G. F. Koster, Phys. Rev. {\bf 94}, 1498 (1954).

\bibitem{Kresse1996} G. Kresse and J. Furthm$\ddot{u}$ller, Phys. Rev. B {\bf 54}, 11169-11186 (1996).

\bibitem{winkler_spin-orbit_2003} R. Winkler, {\it Spin-Orbit Coupling Effects in Two-Dimensional Electron and Hole Systems}, 1st ed. (Springer, Berlin, 2003).

\bibitem{Sabater2013} C. Sabater, D. Gos$\acute{a}$lbez-Mart$\acute{i}$nez, J. Fern$\acute{a}$ndez-Rossier, J. G. Rodrigo, C. Untiedt, and J. J. Palacios, Phys. Rev. Lett. {\bf 110}, 176802 (2013).

\bibitem{graphane} J. O. Sofo, A. S. Chaudhari, and G. D. Barber, Phys. Rev. B {\bf 75}, 153401 (2007); D. C. Elias, R. R. Nair, T. M. G. Mohiuddin, S. V. Morozov, P. Blake, M. P. Halsall, A. C. Ferrari, D. W. Boukhvalov, M. I. Katsnelson, A. K. Geim, K. S. Novoselov, Science {\bf 323}, 610 (2009); J. Zhou, Q. Wang, Q. Sun, X. S. Chen, Y. Kawazoe, and P. Jena, Nano Lett. {\bf 9}, 3867 (2009); J. T. Robinson, J. S. Burgess, C. E. Junkermeier, S. C. Badescu, T. L. Reinecke, F. K. Perkins, M. K. Zalalutdniov, J. W. Baldwin, J. C. Culbertson, P. E. Sheehan, and E. S. Snow, ibid. {\bf 10}, 3001 (2010).

\bibitem{Zhang2014} G.-F. Zhang, Y. Li, and C. Wu, arXiv:cond-mat/1403.0563.




\end{references}

\end{document}